\pdfoutput=1
\documentclass[doc]{apa}
\usepackage{setspace}
\usepackage{graphicx}
\usepackage{latexsym}
\usepackage{subfigure}
\newcommand{\bi}{\begin{itemize}} 
\newcommand{\ei}{\end{itemize}}

\title{On the relationship between the structural and \\ socioacademic communities of a coauthorship network\footnote{Rodriguez, M.A., Pepe, A., ÒOn the Relationship Between the Structural and Socioacademic Communities of a Coauthorship NetworkÓ, Journal of Informetrics, volume 2, issue 3, pages 195-201, ISSN: 1751-1577, doi:10.1016/j.joi.2008.04.002, Elsevier, LA-UR-07-8339, June 2008.}}

\twoauthors{Marko A. Rodriguez\footnote{Both authors contributed equally to this article}}
{Alberto Pepe}

\twoaffiliations{T-7, Center for Non-Linear Studies \\ 
Los Alamos National Laboratory \\
\texttt{marko@lanl.gov}}
{Center for Embedded Networked Sensing \\ 
University of California, Los Angeles \\
\texttt{apepe@ucla.edu}}

\date{\today}
 
\abstract{This article presents a study that compares detected structural communities in a coauthorship network to the socioacademic characteristics of the scholars that compose the network. The coauthorship network was created from the bibliographic record of a multi-institution, interdisciplinary research group focused on the study of sensor networks and wireless communication. Four different community detection algorithms were employed to assign a structural community to each scholar in the network: leading eigenvector, walktrap, edge betweenness and spinglass. Socioacademic characteristics were gathered from the scholars and include such information as their academic department, academic affiliation, country of origin, and academic position. A Pearson's $\chi^2$ test, with a simulated Monte Carlo, revealed that structural communities best represent groupings of individuals working in the same academic department and at the same institution. A generalization of this result suggests that, even in interdisciplinary, multi-institutional research groups, coauthorship is primarily driven by departmental and institutional affiliation.}

\rightheader{Structural and socioacademic communities of a network}
\leftheader{Structural and socioacademic communities of a network}

\begin{document}
\maketitle


\newpage
\section{Introduction}
Scholarly and research environments have progressively become more diversified and interdisciplinary in nature. In interdisciplinary research groups, scholars with different research interests and backgrounds collaborate on different topics and reconcile their research work into joint endeavors \cite{qin}. This type of interaction is generally seen as being driven by the need for a cross-fertilization of ideas when collaborating on topics that bridge across disciplines \cite{niles:inter1975,collab:beaver1978}. The mutual, direct engagement among previously uncorrelated research topics has advantages not only for the researchers, that are able to draw from a wider, diverse intellectual environment, but also for the nature of research performed, that is circulated, validated and enriched by contact with new research and social circles \cite{pierce}. 

Large, interdisciplinary, multi-institutional research centers are interesting environments to study collaboration. The ensemble of social, academic and demographic characteristics found in these centers have the potential to considerably affect collaboration patterns. Research interests \cite{collab:havemann2006} and academic domain \cite{moody} are examples of such characteristics that have been reported to shape the way by which individuals collaborate in a research environment. Coauthorship is a prominent indicator of collaboration in scholarship. However, a number of other indicators of collaboration have been identified and extensively studied and reviewed in the literature. Some of these indicators involve formal, recorded communication and are mostly analyzed by the use of bibliometric methods; besides coauthorship, these include broad categories of research in citation, co-citation and acknowledgment networks (see \citeA{furner} for an extensive review). Other indicators involve less tangible, informal forms of collaboration, such as electronic communication and physical proximity \cite{proximity:katz1994,olson,finholt,pao}.

Coauthorship is the major focus of the work presented in this paper. Relevant to this work are a number of studies that employ network analysis to study coauthorship patterns in academic and scientific circles. \citeA{borner}, for example, posit a weighted graph approach to identify the local and global properties of a scientific coauthorship network to document the emergence of a novel field of science (information visualization). Other domain-specific studies have mined bibliographic databases in the fields of genetic programming \cite{tomassini}, library science \cite{liu} and neuroscience \cite{braun} performing comparative analyses with other scientific collaboration networks. \citeA{lorigo} analyze a large document set of scholarly articles in high energy physics and, rather than inspecting the macroscopic changes in collaboration patterns of this domain, they focus on the impact of Internet-based collaborative technologies on the evolution of remote collaborations. Cross-domain comparative analyses are presented by \citeA{newman_pnas}, who analyzes large databases of papers in the fields of physics, biology, and mathematics exploring social and normative domain differences of coauthorship behavior. All these network-based analyses have proved viable for a number of visualization studies that employ graphical representations of coauthorship networks to uncover macroscopic patterns that network analysis alone might fail to reveal \cite{borner_vis,pubnet}.

In this paper, communities of coauthors in a multi-institutional interdisciplinary research center are uncovered via the use of community detection algorithms. These communities, called \textit{structural communities} (or community structures), are made salient solely by the topological characteristics of the network being analyzed. Structural communities are ``cliquish'' subgraphs composed by groups of vertices that are highly connected between them, but poorly connected to other vertices \cite{girvan-2002}. The study of community structure in networks is particularly important because communities might display local properties that differ greatly from the properties of the network as a whole. Even a very detailed analysis of a network at a global level might fail to uncover specific patterns and characteristics that only exist within tight-knit communities and sub-communities of the network. Community detection algorithms have been previously applied to social networks to uncover the relationship between nationality and collaboration \cite{lozano}, latent communities in large organizations \cite{tyler2003}, political and organizational structures \cite{newman-house}, and to identify communities in networks of collaborating musicians \cite{jazz,rappers}. 

The aim of this article is to detect structural communities of coauthorship and compare them to \textit{socioacademic communities} --- the social and academic groupings of their constituent members. This comparative analysis uncovers the specific socioacademic characteristics that are best described by community structures in a coauthorship network. Four different community structure configurations are found, using the following community detection algorithms: leading eigenvector \cite{newman-eigen}, walktrap \cite{latapy}, edge betweenness  \cite{girvan-2002} and spinglass \cite{spinglass:reichardt2006}. Subsequently, the community structures detected via the four different algorithms are subjected to Pearson's $\chi^2$ test for statistical independence \cite{parame:sheskin2004} to determine whether the latent structural communities in the research group's coauthorship network are dependent (or independent) of the various socioacademic characteristics of the scholars.

\section{Socioacademic and structural communities\label{sec:summary}}
Community structure and many other social network indices are largely based on the topology of the network, rather than specific social, demographic or academic characteristics of the individuals in the network (i.e.~socioacademic characteristics). In the research group analyzed in this study, scholars from different institutions and departments collaborated together to produce jointly coauthored papers, books and conference proceedings. 

Community detection algorithms are useful at capturing the structural communities of the resulting coauthorship network. However, as coauthorship networks are essentially made up of scholars, the various socioacademic characteristics of these scholars can be regarded as parallel, socioacademic communities. Note that it is not necessary that these various structural and socioacademic communities overlap. For example, two scholars might be members of the same structural community in a coauthorship network, but have different academic affiliations and thus, be members of different institutional communities. However, when two communities do overlap in a non-random manner, much can be said about the semantics of the latent structural community. Before discussing the relationship between structural and socioacademic communities, this section will review the various types of communities present in the research group under study.

\subsection{Socioacademic communities}
The research center analyzed here is the Center for Embedded Networked Sensing (CENS), a National Science Foundation venture involving scholars at all levels (faculty, scientists, engineers, graduate and undergraduate students) from five member institutions (University of California at Los Angeles, University of Southern California, University of California Riverside, California Institute of Technology, and University of California at Merced). The type of research conducted at CENS spans across a wide spectrum of disciplines and applications: from biology to seismology, from wireless telecommunications to statistics.

The coauthorship network employed in this study was constructed from $560$ manuscripts ($379$ conference papers, $163$ journal articles, $17$ book chapters and $1$ book) published over a ten year period ($1998$-$2007$) by CENS scholars and available at the the CDL eScholarship repository (http://repositories.cdlib.org/cens). In particular, only those papers were considered that have at least one of the authors affiliated with CENS at the time of publication of the paper. Ultimately, the generated coauthorship network contained 291 vertices (scholars) and 2536 edges (coauthoring events). For every scholar in the network, details about the following socioacademic characteristics were gathered: academic department, academic affiliation, country of origin, and  academic position. Table \ref{tab:stats} presents the frequency distribution of the values for the aforementioned socioacademic characteristics.

\begin{table*}[h!]
\centering
  \begin{minipage}{6cm}
\begin{footnotesize}
\begin{tabular}{ll}
\hline
\textbf{count}&\textbf{department}\\
\thickline
113&Computer Science\\
80&Electrical Engineering\\
23&Civil Engineering\\
19&Biology\\
9&Information Science\\
7&Environment\\
5&Education\\
4&Marine Biology\\
\hline
\end{tabular}
\end{footnotesize}
  \end{minipage}
  \begin{minipage}{6cm}
\begin{footnotesize}
\begin{tabular}{ll}
\hline
\textbf{count}&\textbf{affiliation}\\
\thickline
148&UCLA\\
66&USC\\
10&MIT\\
8&Caltech\\
7&UC Riverside\\
7&UC Berkeley\\
4&UC Merced\\
3&SUNY Stony Brook\\
\hline
\end{tabular}
\end{footnotesize}
  \end{minipage}
  \begin{minipage}{5.5cm}
\begin{footnotesize}
\begin{tabular}{ll}
\hline
\textbf{count}&\textbf{origin}\\
\thickline
120&United States     \\
33&India\\
24&China\\
10&Italy\\
 9&South Korea\\
 5&Australia\\
 4&Greece\\
 3&Iran\\
\hline
\end{tabular}
\end{footnotesize}
  \end{minipage}
  \begin{minipage}{7.3cm}
\begin{footnotesize}
\begin{tabular}{ll}
\hline
\textbf{count}&\textbf{position}\\
\thickline
97&PhD student\\
67&Other researcher\\
44&Professor\\
21&Postdoc\\
21&Associate Professor\\
20&Assistant Professor\\
5&Undergraduate Student\\
3&Lecturer\\
\hline
\end{tabular}
\end{footnotesize}
  \end{minipage}
\centering
\caption{\label{tab:stats} Frequency counts (top 8) for the socioacademic characteristics of the scholars under study: academic department, academic affiliation, country of origin, and academic position.}
\end{table*}

\subsection{Structural communities}
Structural communities relative to the coauthorship network studied here were identified using four different community detection algorithms: 
\begin{seriate}
\item \textit{leading eigenvector}, a recent and popular technique, based on the definition of the modularity function in terms of the eigenspectrum of matrices \cite{newman-eigen},
\item \textit{walktrap}, a technique based on random walks \cite{latapy},
\item \textit{edge betweenness}, the earliest community detection technique, based on vertex betweenness centrality \cite{girvan-2002} 
\item \textit{spinglass}, a technique based on a spin-glass model and simulated annealing \cite{spinglass:reichardt2006}. 
\end{seriate}

\begin{figure*}[h!]
\label{fig:communities}
\centering
\includegraphics[width=1\textwidth]{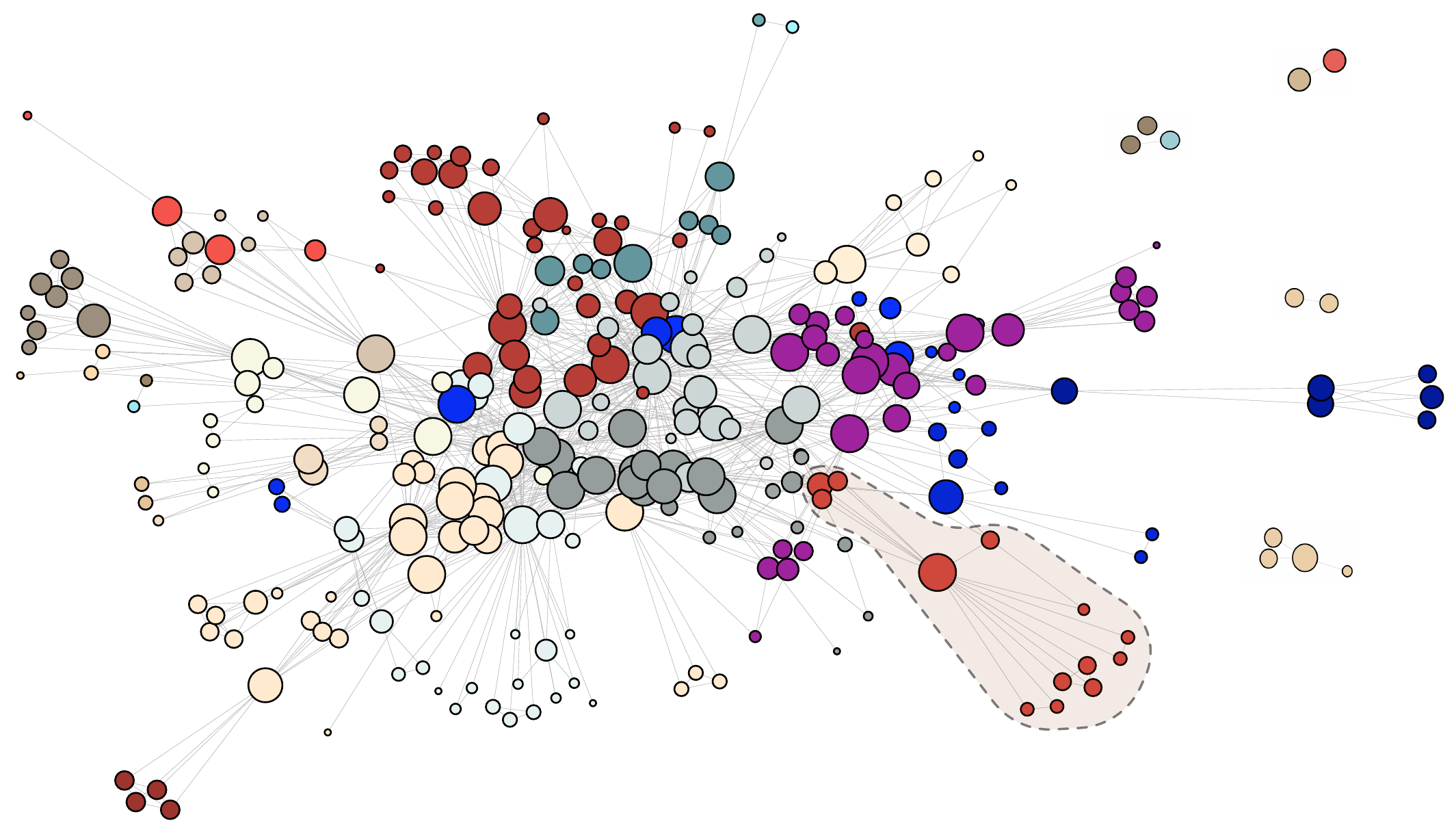}
\caption{\label{fig:network}Detected structural communities in the coauthorship network under study detected according to the leading eigenvector algorithm. Each community is represented using a different color (or shade). Vertex diameter represents the eigenvector centrality score of the vertex, where more central vertices have larger diameters. The community highlighted in the lower right corner is further analyzed in Figure \ref{fig:subgraph}.}
\end{figure*}

The structural communities found in the coauthorship network via the leading eigenvector algorithm are diagrammed in Figure \ref{fig:network} according to the Fruchterman-Reingold network layout algorithm \cite{layout:fruchter1991}. The leading eigenvector community detection algorithm deconstructs the coauthorship network into $27$ structural communities and it assigns a membership value to each node. Note that the membership value is a nominal value identifying distinction, not relative similarity between identified communities. Thus, scholars that are in the same structural community are given the same membership value. With respect to Figure \ref{fig:network}, the membership value is identified by a unique color (or shade). The diameter of the vertices in the figure represents the weighted eigenvector centrality score of the vertices, where more central vertices have larger diameters \cite{power:bonacich1987}. Finally, note that the community highlighted in the lower right portion of the figure is further analyzed in the next section.

\begin{figure*}[h!]
\centering
\includegraphics[width=0.65\textwidth]{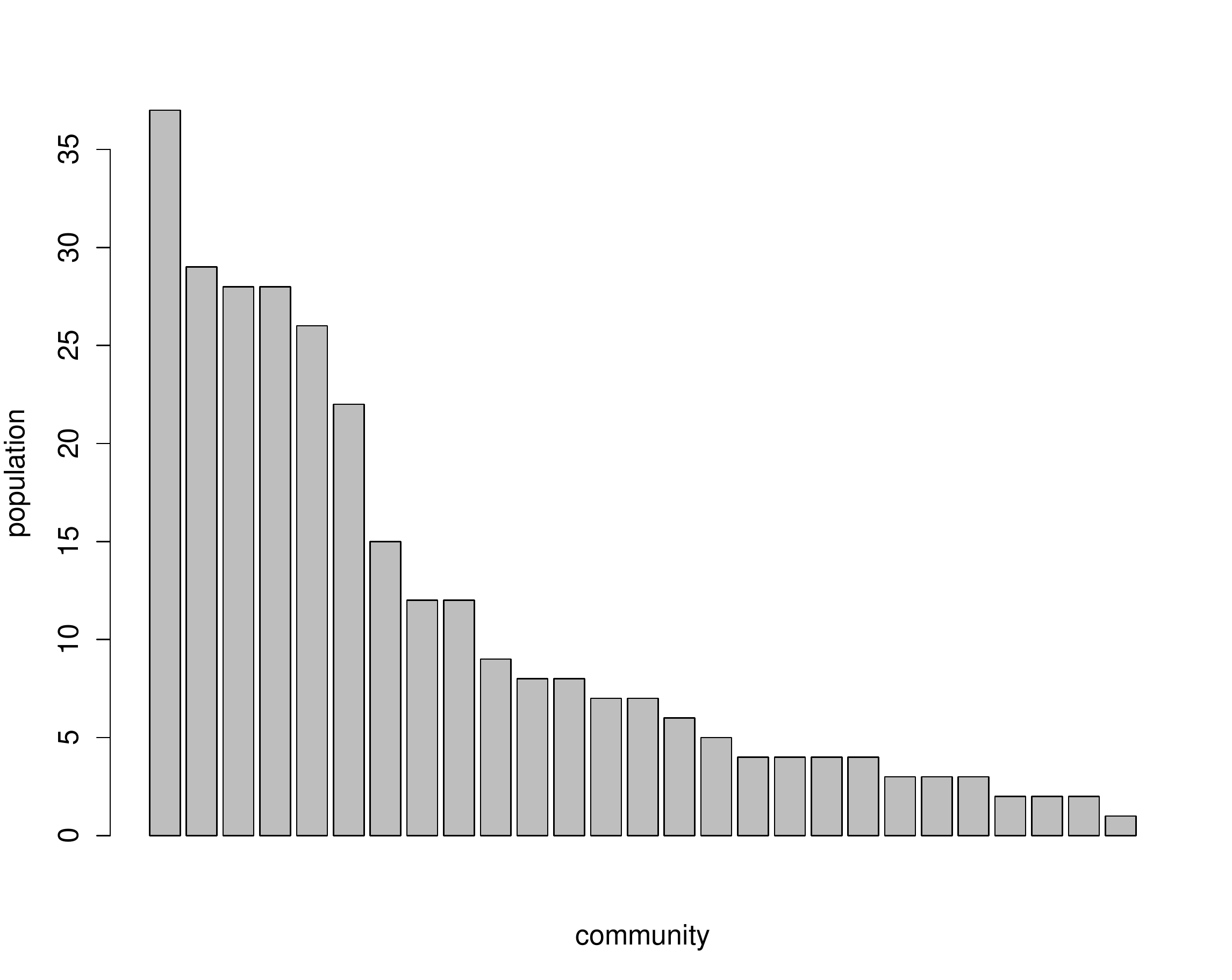}
\caption{\label{fig:community-frequency}The number of scholars in the $27$ identified structural communities according to the leading eigenvector community detection algorithm.}
\end{figure*}

Some fundamental statistics relative to the coauthorship network studied here are presented in Table \ref{tab:netstats}. The coauthorship network is found to be highly connected, with five connected components (distinctly separated subnetworks) with the vast majority of the vertices being within the largest component (280 out of a total of 291 vertices). This means that besides the four small separate components, the interdisciplinary research group studied here is perceived, as a whole, as a single coauthoring community. Figure \ref{fig:community-frequency} presents the number of scholars identified in the $27$ structural communities identified by the leading eigenvector community detection algorithm.

\begin{table}[h!]
\label{table:net-stats}
\begin{footnotesize}
\begin{tabular}{ll}
\hline
\textbf{variable}&\textbf{value}\\
\thickline
vertices&291\\
edges&2536\\
cliques&14\\
connected components&5 (280, 4, 3, 2, 2)\\
connectedness&0.926\\
$\gamma$ coefficient&1.460\\
clustering coefficient&0.330\\
\hline
\end{tabular}
\caption{\label{tab:netstats}A summary of some fundamental statistics for the coauthorship network under study.}
\end{footnotesize}
\end{table}

In this network, like most natural networks, there are few highly connected vertices and many lowly connected vertices. This is further made salient by the eigenvector centrality scores represented by the diameter of the vertices in Figure \ref{fig:network}. The clustering coefficient value found for the coauthorship network presented here (0.33) is similar to that found in the coauthorship networks in the field of mathematics (0.34), but much lower than what is generally found in the physical sciences and biology \cite{newman-siam}. This suggests less-cliquish, sparse collaboration patterns among scholars and is perhaps indicative of an interdisciplinary community that is more fragmented in their research agenda.

\section{Results\label{sec:results}}
In order to test for independence between the detected structural communities and the explicit socioacademic communities of the coauthorship network under study, a Pearson's $\chi^2$ (i.e.~chi-squared) analysis was conducted. For each one of the four socioacademic characteristics (academic department, academic affiliation, country of origin and academic position), four contingency tables were created, one for each community detection algorithm employed (leading eigenvector, walktrap, edge betweenness, spinglass\footnote{The spinglass community detection algorithm requires a connected network, therefore the analysis was performed only on the the largest connected component which contained 280 of the total 291 nodes. Thus, 11 nodes were left out of this analysis.}). In each of these tables, the $x$-axis elements represent a distinct community membership value as identified by the specific community detection algorithm used and the $y$-axis represents each distinct value for a particular socioacademic characteristic. Cell values in each contingency table identify the number of observed occurrences of an $x$/$y$ relationship.

The contingency tables were then subjected to a Pearson's $\chi^2$ test with a simulated Monte Carlo to determine whether each socioacademic property was dependent or independent of the assigned community membership value. The results of the $\chi^2$ tests for each socioacademic characteristic are presented in Table \ref{tab:chisquare}, where the $p$-value denotes the probability that an association is random (i.e.~a $p$-value greater than $0.05$ is generally considered statistically independent). The $p$-values obtained via the four community detection algorithms all demonstrate that both scholar's department and affiliation are dependent on the identified structural community of the scholar. On the other side of the spectrum, the academic position and country of origin of a scholar are independent of the identified structural community of the scholar. This latter result was disconfirmed only by the spinglass community detection algorithm that found academic position to be statistically significant.  

It is important to note that the dataset under study lacks socioacademic values for some of the individuals in the coauthorship network; the number of \textit{null} values for each socioacademic characteristic are also displayed in Table \ref{tab:chisquare}. The absence of these values, coupled with the size and configuration of the network, resulted in some of the contingency tables being sparsely populated (i.e.~a high proportion of the cell expected values being equal to 0 or 1). Table sparseness is a cause of concern for the validity of the $\chi^2$ test for independence. In order to remedy this situation, a Monte Carlo exact test was conducted. This method has been recognized as a solution to table sparseness for categorical multivariate data analysis \cite{agresti,reiser}.

\begin{table}[h!]
\label{tab:chisquare}
\begin{footnotesize}
\begin{tabular}{lccccc}
\hline
\textbf{socioacademic characteristic}&\textbf{null values}&\textbf{LEV}&\textbf{WT}&\textbf{BC}&\textbf{SG}\\
\thickline
Academic Department&5&$0.0001$&$0.0005$&$0.0005$&$0.0005$\\
Academic Affiliation&0&$0.0264$&$0.0007$&$0.0265$&$0.0005$\\
Country of Origin&52&$0.2403$&$0.4130$&$0.2293$&$0.0934$\\
Academic Position&13&$0.7166$&$0.1486$&$0.1672$&$0.0453$\\
\hline
\end{tabular}
\caption{\label{tab:chisquare} $p$-values for the $\chi^2$ tests of the contingency tables (academic department, academic affiliation, country of origin and academic position) obtained via community detection algorithms: leading eigenvector (LEV), walktrap (WT), edge betweenness (EB), spinglass (SG).}
\end{footnotesize}
\end{table}

\begin{figure}[h!]	
	\begin{center}
    \mbox{
\subfigure[Academic Department]{\includegraphics[width=0.45\textwidth]{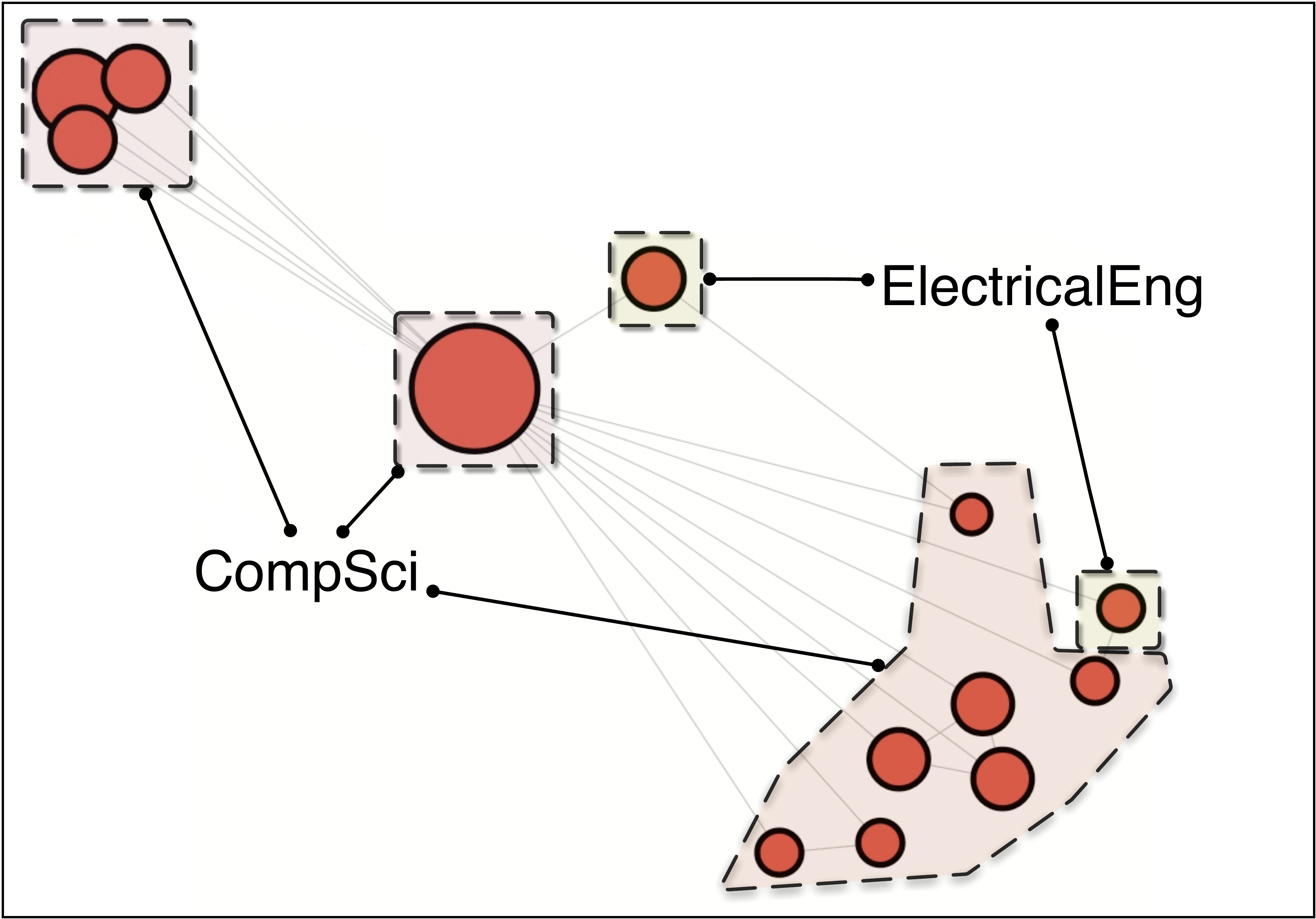}}
\subfigure[Academic Affiliation]{\includegraphics[width=0.45\textwidth]{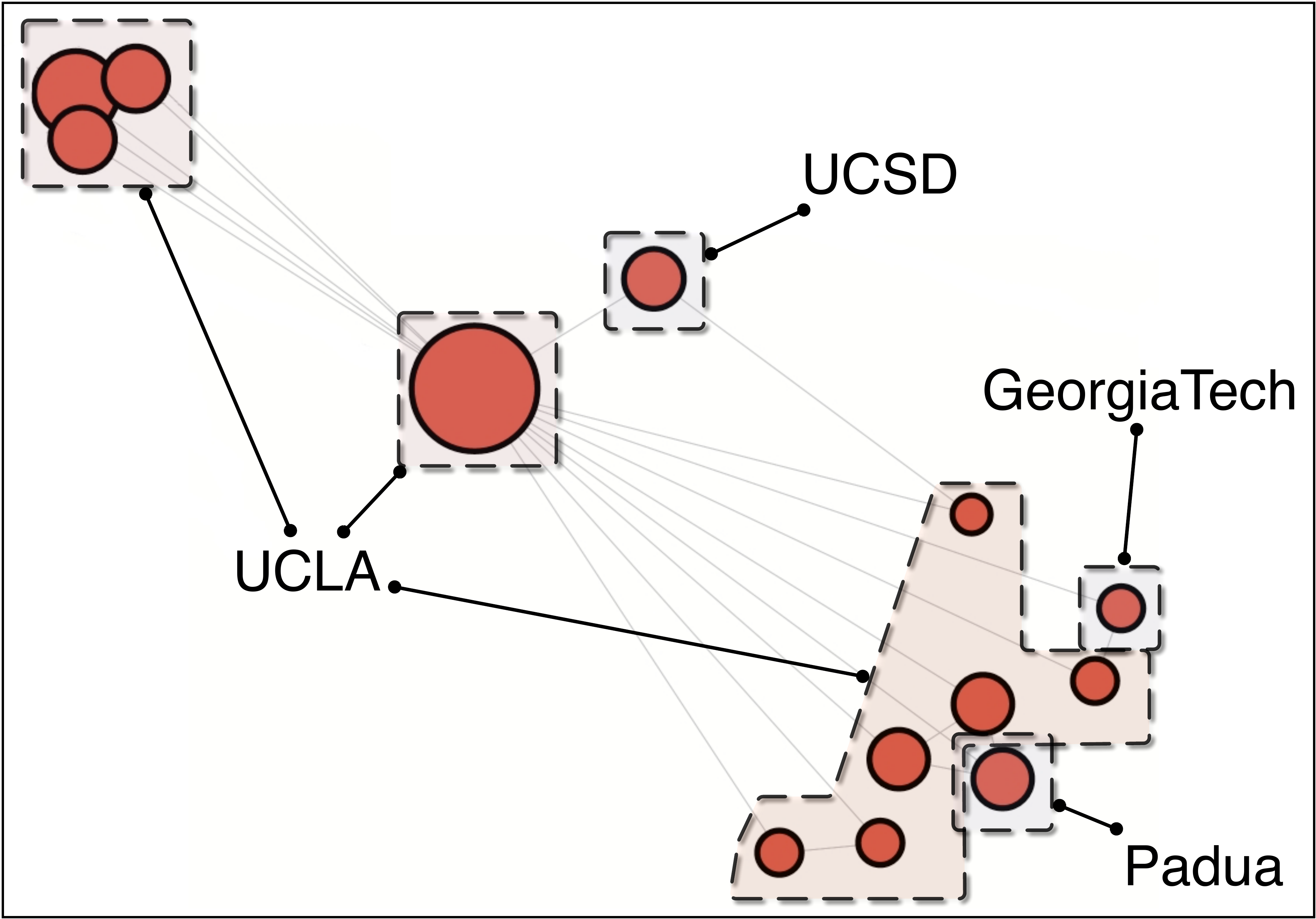}}\quad
}
    \mbox{
\subfigure[Country of Origin]{\includegraphics[width=0.45\textwidth]{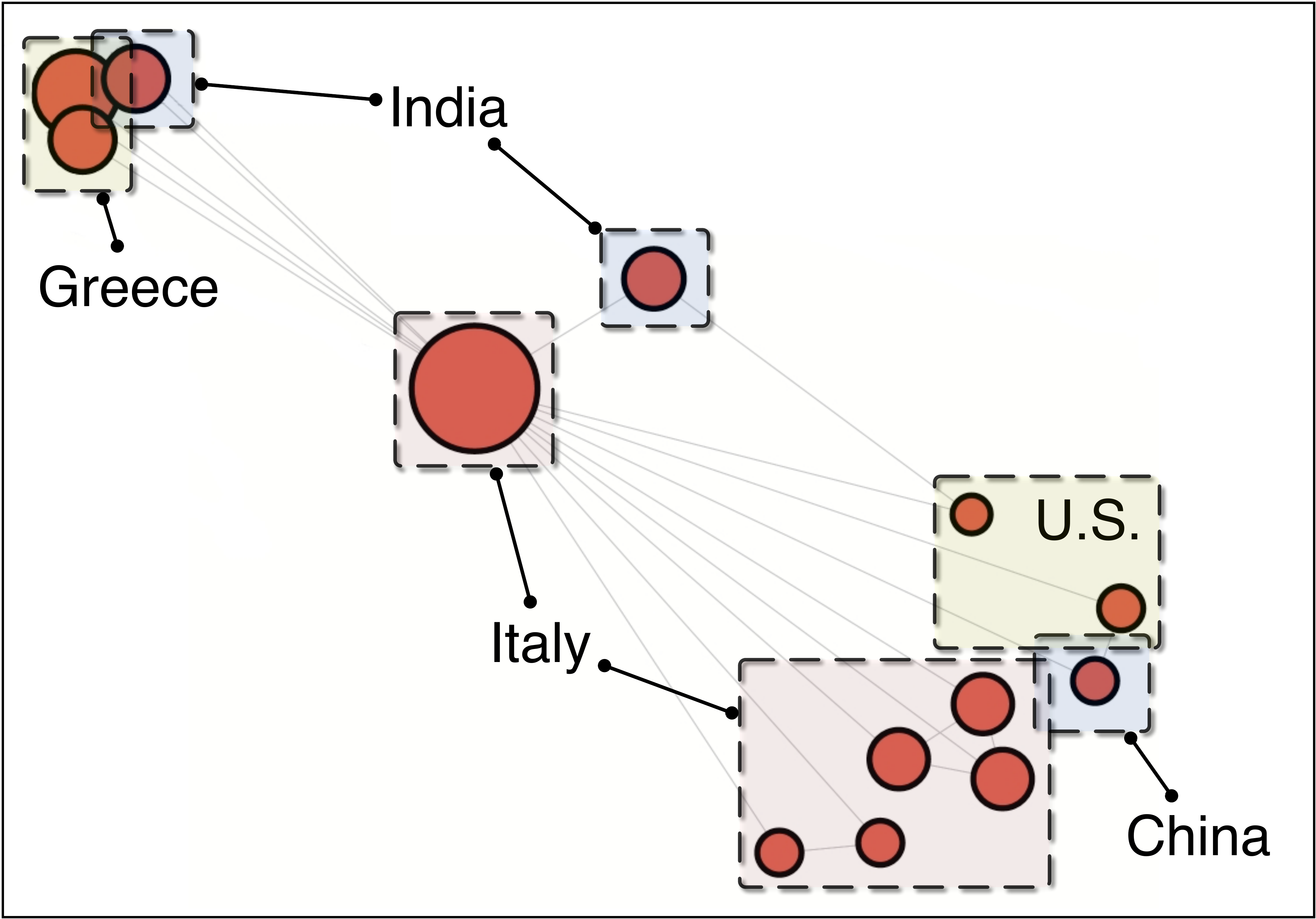}}
\subfigure[Academic Position]{\includegraphics[width=0.45\textwidth]{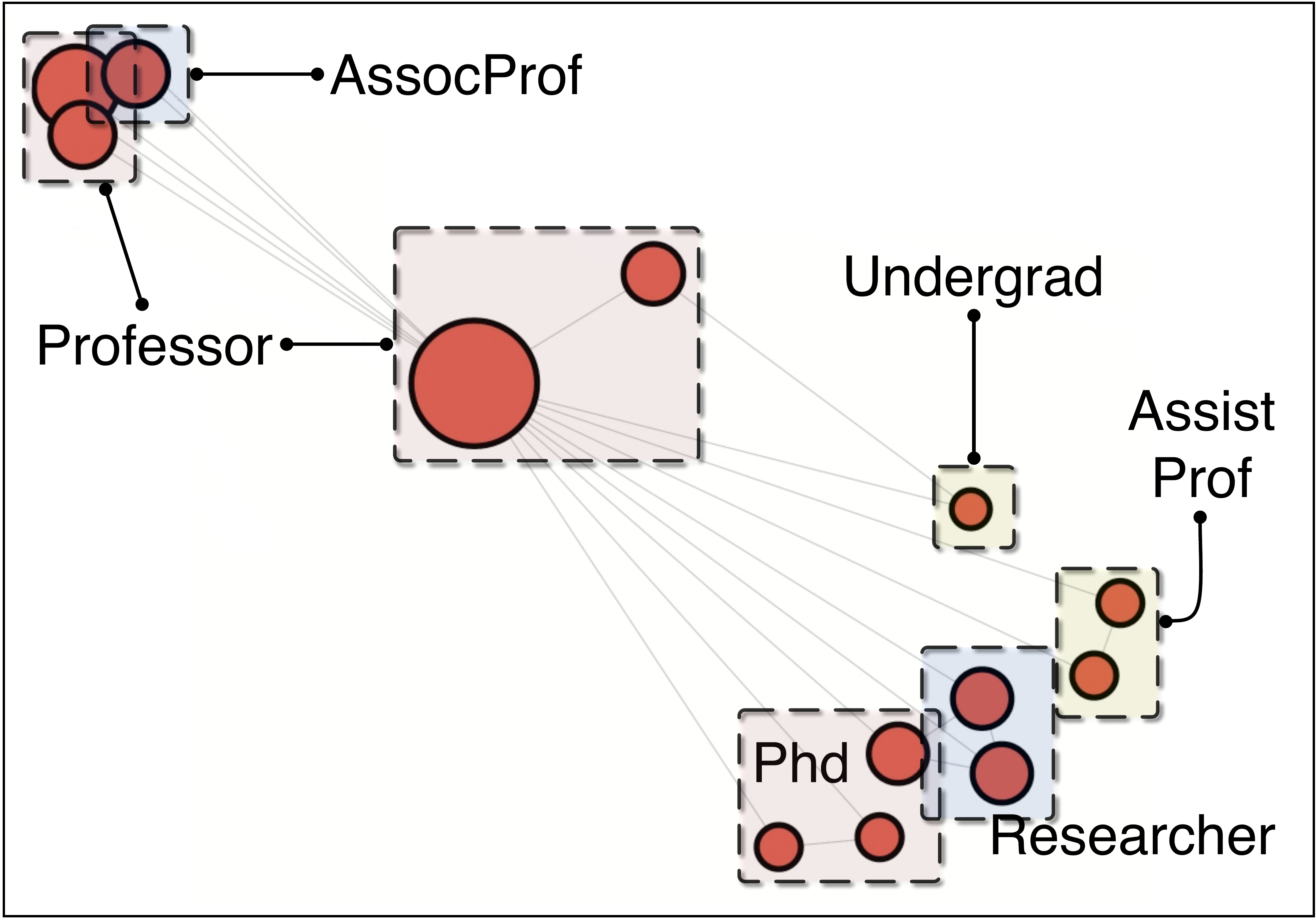}}\quad
}	
\caption{\label{fig:subgraph}Socioacademic characteristics of a specific structural community of the coauthorship network under study. In this example, community structure best represents department and affiliation and poorly represents origin and academic position.}
	\end{center}
\end{figure}

Figure \ref{fig:subgraph} presents an anecdotal example of the aforementioned finding: the structural community highlighted in the lower right portion of Figure \ref{fig:network}. The four socioacademic properties of the scholars have been identified and annotated: department, affiliation, country of origin, and academic position. Returning to the $\chi^2$ results in Table \ref{tab:chisquare}, one would expect to see department and affiliation to match closely to the membership community, and country of origin and academic position to be random. This prediction is confirmed exactly in the visual investigation of the specific community presented in Figure \ref{fig:subgraph}. 

Figure \ref{fig:subgraph}a indicates the academic departments of the scholars in the community: these are almost entirely scholars in computer science, with the exception of two electrical engineers. Academic affiliation is more variegated (Figure \ref{fig:subgraph}b) with three different scholars in the community belonging to three different institutions, yet all other scholars are from UCLA. The other two views of the same community (Figures \ref{fig:subgraph}c and \ref{fig:subgraph}d) present more fragmented scenarios: scholars come from five different countries and six different academic ranks. This visual analysis thus confirms the quantitative findings of the $\chi^2$ test.  

\section{Conclusion\label{sec:conclusion}}

The analysis presented in this article compared communities in the studied coauthorship network identified by the community detection algorithms to communities made explicit in the socioacademic profile of the constituent community members. The structural communities detected and the socioacademic communities were compared using a Pearson's $\chi^2$ test with a simulated Monte Carlo. Results demonstrated that the community structures revealed by four different community detection algorithms (leading eigenvector, walktrap, edge betweenness and spinglass) all successfully capture specific existing socioacademic characteristics of individual scholars, namely academic department and affiliation. On the other hand, they fail to correctly capture demographic characteristics such as the individuals' country of origin. With the exception of the spinglass algorithm, all community detection methods indicate statistical independence for academic position.  

These results might be generalized in several ways. First, results might be used to make specific policy recommendations for the research group under study, the Center for Embedded Networked Sensing (CENS). The analysis has shown that coauthoring community structures overlap with socioacademic communities sharing characteristics such as academic affiliation and department. In other words, the results reveal that most of the coauthorship activity appears within communities of scholars related by the same area of expertise (department) and within the same institution (affiliation). This finding is in contrast with the scope and goals of inter-disciplinary multi-institutional research centers.

Second, the results could be extended to other coauthorship networks. Similar comparative analyses could be performed on coauthorship networks of other scholarly domains to find out how different socioacademic characteristics describe the structural communities of diverse fields and/or research environments. In these studies, a number of additional social, academic and demographic parameters could be added to the socioacademic study, including physical location, first language, and research specialization. 

Finally, these types of studies might prove useful to predict either the topological or socioacademic configuration of a network when either data is scarce. For example, the coauthoring community structures revealed in this study were found to be strongly correlated with academic affiliation and department. Thus, it might be possible to infer the academic affiliation and department of certain community members for which no data is available, simply based on the topological community structure to which they belong. Similarly, unknown coauthoring relationships might be inferred among individuals whom, though not being part of the same community structure, share a set of socioacademic characteristics.  

\section*{Acknowledgments}
The authors would like to thank the developers of iGraph (http://cneurocvs.rmki.kfki.hu/igraph/) for their wonderful suite of network analysis algorithms. Furthermore, the data used in this research was provided by CENS through CDL's eScholarship repository.

\end{document}